\documentclass[aps,preprint]{revtex4}
\usepackage{graphicx}
\begin{document}

\title{Violation of the Universal Behavior of Membranes
Inside Cylindrical Tubes at Nanoscale}

%\maketitle

\author{E.\ Perim$^1$, A.\ F.\ Fonseca$^1$, N.\ M.\ Pugno$^{2,3,4}$, D.\ S.\ Galvao$^1$ }

\affiliation{$^1$Applied Physics Department, State University of Campinas, Campinas-SP, Sao Paulo, Brazil.\\
$^2$Laboratory of Bio-inspired and Graphene Nanomechanics, Department of Civil, Structural and Mechanical Engineering, University of Trento, via Mesiano, 77, 38123 Trento, Italy.\\
$^3$Center for Materials and Microsystems, Fondazione Bruno Kessler, via Sommarive 18, 38123 Povo (Trento), Italy.\\
$^4$School of Engineering and Materials Science, Queen Mary  University of London, Mile End Road, London E1 4NS, UK.}

\date{\today}

\begin{abstract}
  Recently, it was proposed based on classical elasticity theory and
  experiments at macroscale, that the conformations of sheets inside
  cylindrical tubes present a universal behavior. A natural question
  is whether this behavior still holds at nanoscale. Based on
  molecular dynamics simulations and analytical modeling for graphene
  and boron nitride membranes confined inside carbon nanotubes, we
  show that the class of universality observed at macroscale is
  violated at nanoscale. The precise origins of these discrepancies is
  addressed and proven to be related to both surface and atomistic
  effects.
\end{abstract}

\maketitle

%\clearpage

Systems exhibiting classes of universalities (where the main
properties are material and scale independent\cite{ref1}) are of great
interest since it is possible to reliably predict and generalize the
properties of a larger number of structures.

With the advent of nanotechnology, many systems
exhibiting counter-intuitive and unusual behaviors have been reported,
such as auxetic buckypapers~\cite{ref2,bucky}, atomic suspended
chains~\cite{ref3}, and `exotic' metallic structures that can exist
only at nanoscale~\cite{ref4}. Thus, much effort has been devoted to adjust
macroscopic models in order to understand the nanoscale effects and
the origin of these unusual behaviors.

One important example of these approaches is the macroscopic or
continuum modeling of the elastic properties of graphene~\cite{ref5}
(single layer graphite). Nevertheless, it was demonstrated~\cite{ref6} that
continuum models fail to describe the detailed elastic behavior of
single-layer graphene, although they can reliably describe the
properties of many-layers systems.

\begin{figure}[ht]
  \begin{center}
\includegraphics[width=50mm]{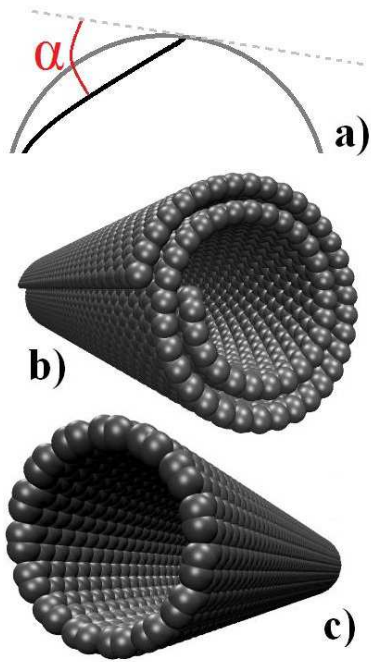}
  \end{center} 
  \caption{(Color online) (a) the definition of the $\alpha$ angle.
    (b) and (c), carbon nanoscroll and carbon nanotube structures,
    respectively. See text for discussions.}
  \label{fig1}
\end{figure}

Recently, Romero and co-workers~\cite{ref7}, after here named RWC
model, published a detailed work on the morphology of coiled elastic
sheets inside cylinders. The RWC model is based on classical continuum
mechanics and proposes that the observed elastic conformations of
the sheets inside the cylinders should exhibit an universal behavior which would be expressed by the $\alpha$ angle
formed between the sheets and the tubes (Figure
\ref{fig1}{(a)}). These $\alpha$ angle values
should be the same, independently on the size or type of material.
These predictions were validated by a series of experimental
tests~\cite{ref7}.

However, as the RWC model was derived for
macroscale systems, it neglects some aspects, which become very
important at nanoscale, for instance, the significant stickiness generated by
the van der Waals (vdW) forces. Another important aspect not
considered in the RWC model was the tube topology aspects at atomistic scale
(such as, tube chirality).

In the light of the more recent work of Zhang and
collaborators~\cite{ref6}, a natural question is whether this
so-called ``universal'' behavior observed at macroscale would hold at
nanoscale. In order to address this important question we have carried
out fully atomistic molecular dynamics (MD) simulations for nanoscale
structural models in association with analytical modeling.

As the RWC model investigated folded sheets placed inside a cylindrical
tube, we used graphene (G)~\cite{ref8} and boron nitride (BN)~\cite{ref9} membranes
and carbon nanotubes (CNTs)~\cite{ref10} as corresponding nanostructures for sheets and tubes, respectively. The
concentrically cylindrical folded (rolled up) G and BN membranes
inside the CNTs~\cite{nestedscrolls} generate the so-called
nanoscrolls~\cite{origami,scrollrev} (Figure \ref{fig1}{b)}).

As shown in Figure \ref{fig1}, nanoscrolls are papyrus-like seamlessly
wrapped sheets with open ends and can exhibit (depending on the
diameter and the materials they are composed of) higher structural
stability than their corresponding planar-layered
conformations~\cite{scrollosc}, resulting from the attractive vdW
forces, which can overcome the elastic ones~\cite{ref11,ref12}.
Carbon~\cite{ref13} and BN~\cite{ref14,ref15} nanoscrolls (CNSs and
BNNSs, respectively) have been already experimentally realized. For
these nanostructures the vdW forces are of major importance in
defining their structural stability and they cannot be neglected as
they (in general) are in macroscopic models. In the present work, our
models consisted of either carbon or BN membranes coiled/scrolled
inside CNTs. Without lack of generality, all G and BN membranes, as
well as the used CNTs were of zigzag type~\cite{ref11}.  We have
considered scrolls formed from membranes with rectangular dimensions
of (from 160 up to 320 \AA) by 32 \AA\ which are rolled up as archimedian spirals around the axis along the direction of their smaller side, with constant layer separation of 3.4 \AA\ and inner diameter approximately of 20 \AA ~\cite{ref11}. The CNT diameter and length
were chosen to be compatible with the scroll dimensions, considering
the cases of uncompleted rolling up to many-layers scrolls.

In order to directly contrast our results with the ones from the RWC
model (where the tube dimensions do not change), our CNTs were kept
frozen in all the simulations.

All MD simulations were carried out using the universal molecular
force field (UFF)~\cite{ref16}, as implemented in the Materials Studio
suite~\cite{ref17}. UFF is a well-known and tested
force field and includes bond stretch, bond angle bending, inversion,
torsions, rotations and vdW terms. The MD simulations were carried out
within the NVE (number of atoms, volume and energy constant) ensemble
with convergence criteria of $10^{-5}$ kcal/mol for energy and 5
$\times 10^{-3}$ kcal \AA/mol for maximum force among atoms,
respectively.  No explicit charges were used and all atoms were
assumed as having partial double bonds and sp$^2$ hybridizations. This
approach has been proven to be very effective in the description of
mechanical and structural properties of CNTs and
scrolls~\cite{ref11,ref12}.

In order to test the ``universal'' behavior of the RWC model at
nanoscale, we analyzed its critical variable prediction: the angle $\alpha$
formed by the sheets with respect to the tube wall (Figure
\ref{fig1}{a)}). From the RWC model this angle is
supposed to have a universal value of $24.1^{\circ}$.

There are many possible configurations for the combination of scrolls
and tubes, but we restricted ourselves, due to the lack of space here, to
two major cases; (I) the one at which the tube has a diameter large
enough for the sheets inside the tubes retain their scrolled
conformation and;(II) the case where the sheets lengths are smaller
than the tube diameter.

We will start discussing the case (I). In order to preserve their
structural stability isolated scrolls must have an inner diameter
around 20 \AA~\cite{ref11,ref12}. This scenario is analogous to the
one treated by the RWC model for $0.26 < D/L < 0.32$, where $D$ is the
tube diameter and $L$ the scroll length.

\begin{figure}[ht]
  \begin{center}
    \includegraphics[width=65mm]{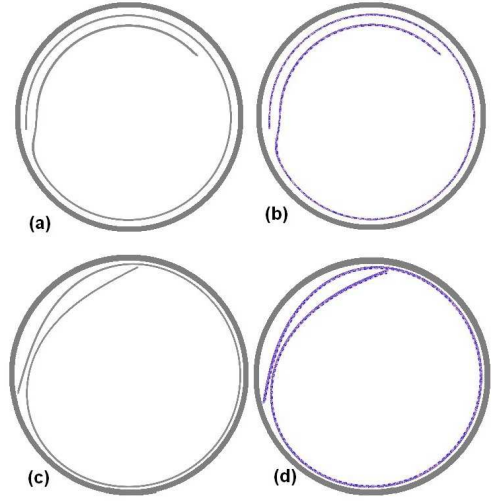}
  \end{center} 
  \caption{(Color online) Optimized (a) CNS and (b) BNNS inside a CNT
    with full vdW interactions; (c) CNS and (d) BNNS with scaled down vdW ones. The measured $\alpha$ angles for cases (c) and
    (d) were both equal to 28$^{\circ}$. See text for discussions.}
  \label{fig2}
\end{figure} 

In Figure \ref{fig2} we present the obtained scroll morphologies for
CNSs and BNNSs, respectively. Accordingly to the RWC model, the scrolls
should exhibit two detached regions, for both inner and outer scroll
layers but, as we can see from Figure {\ref{fig2}{(a)} and \ref{fig2}{(b)},
this did not occur, either to carbon or BN scrolls, showing that
the so-called ``universal'' behavior predicted by the RWC model is
violated at nanoscale, and, consequently, there is no
real ``universal'' behavior for sheets confined inside cylindrical
tubes.

For the case (II), where we have sheets that are not larger enough to form complete scrolled structures, we found similar results (see Figure {\ref{fig3}{(a)} and \ref{fig3}{(b)}). No detachment of the sheets from the tubes were observed and $\alpha$ angle values are again different than the ones predicted by the RWC model.

\begin{figure}[ht]
  \begin{center}
    \includegraphics[width=65mm]{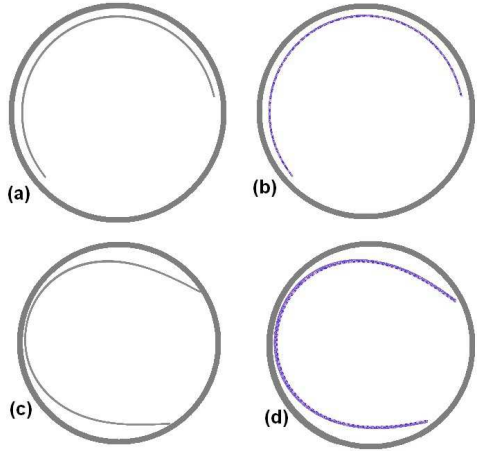}
  \end{center} 
  \caption{(Color online) Optimized (a) Unformed scrolled CNS and (b) BNNS inside a CNT
    with full vdW interactions; (c) CNS and (d) BNNS with scaled down vdW ones. The measured $\alpha$ angles for cases (c) and
    (d) were both equal to 28$^{\circ}$. See text for discussions.}
  \label{fig3}
\end{figure} 

It remains to be elucidated the origin of these apparent
discrepancies. As above mentioned, one essential aspect of the models
at nano and macro scale are the relative importance of the vdW forces,
which were not explicitly taken into account in the RWC model.

In order to test whether the vdW forces would be in the origin of
these discrepancies, we modified the terms related to the vdW forces
in our molecular force field and reran the MD simulations. We
considered different cases, where the vdW interactions were
gradually decreased until the limit situations where the vdW forces
were mainly of repulsive type. This can be done by changing the parameter
values in the force field which control the well-depth energy of the
Lennard-Jones potential associated with the vdW forces. In our MD
simulations these parameters were gradually decreased up to seven
orders of magnitude with relation to their standard values. If the vdW
forces are the only responsible cause for the $\alpha$ discrepancies,
it should be expected that decreasing the vdW forces would make the
$\alpha$ values to converge to the macroscopic predicted values.

\begin{figure}[ht]
  \begin{center}
    \includegraphics[height=126mm,width=140mm,clip]{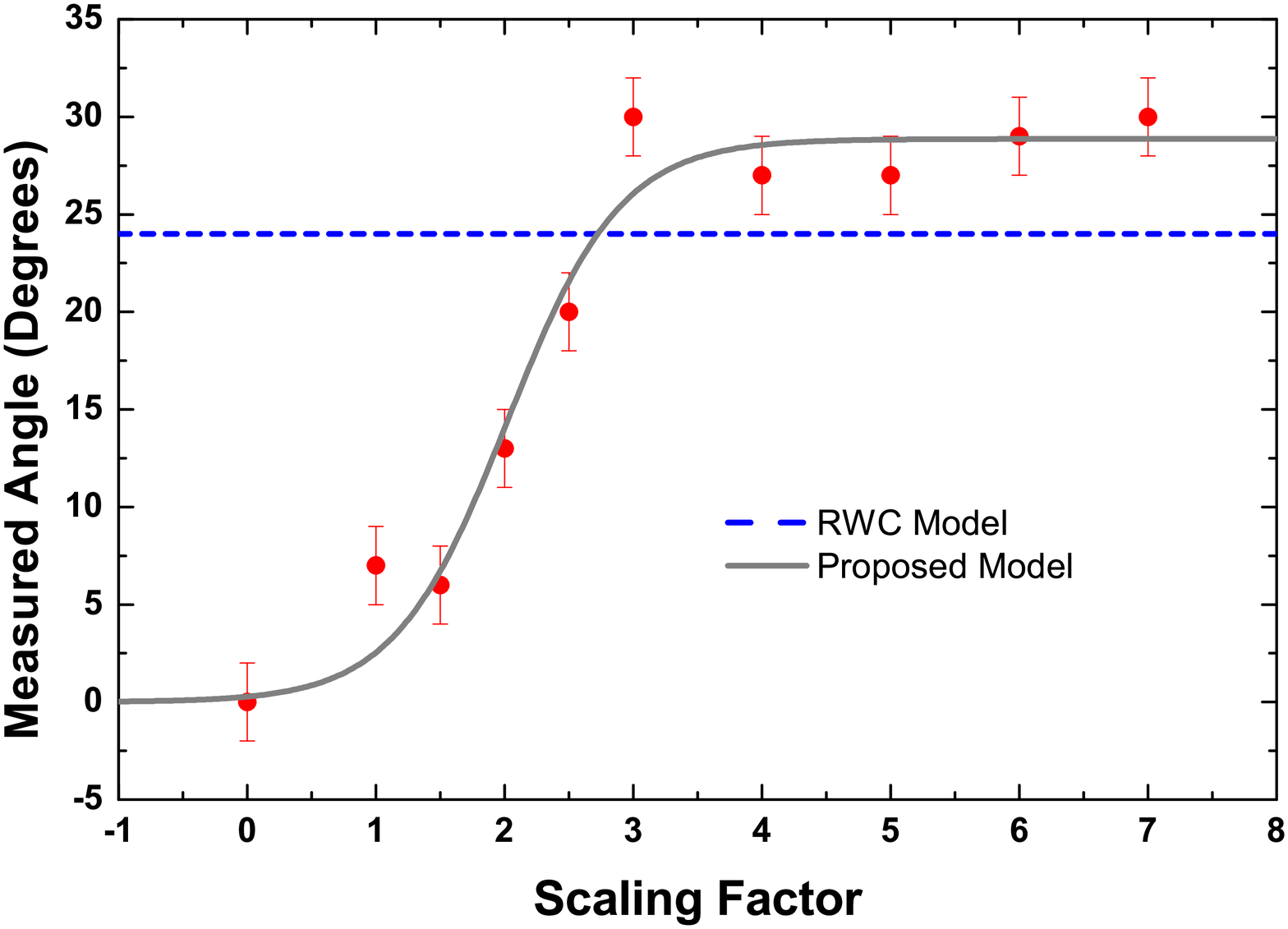}
  \end{center} 
  \caption{(Color online) Measure $\alpha$ angle values as a function
    of the scaled down van der Waals terms. Results obtained for
    carbon membranes. The Lennard-Jonnes potential well depth
    $\epsilon(n)$, proportional to the surface energy $\gamma$, is
    scaled as $\epsilon(n)=\epsilon_0/10^{n}$, where $n$ is the
    scaling factor. See text for discussions.}
  \label{fig4}
\end{figure} 

In Figure \ref{fig4} we present the results for the carbon scrolls;
similar results were obtained for the BN ones. As we can see from the
Figure, scaling down the vdW interactions indeed make the $\alpha$
values become closer to the ``universal'' value predicted by the RWC
model. However, these $\alpha$ values vary and, more importantly, they
do not converge to the expected macroscopic value.  Thus, the vdW
contributions alone cannot be associated with the $\alpha$
discrepancies. One possibility is that these discrepancies have a pure
atomistic origin, as recently proposed in the literature~\cite{ref6}.

In order to test this hypothesis, we have tried a different approach by including a simple modification in the RWC model for the contact
angle of the sheets inside the tubes.

We assume that the variation of the $\alpha$ angle with respect to the
its macroscopic ``universal'' value is due to the presence of a
pressure variation imposed by the adhesion and a bending stiffness
variation due to atomistic effects of a single atomic
layer~\cite{ref6}.

From the classical elastic solution~\cite{ref19}, the pressure at the
contact tip is:
\begin{equation}
\label{press}
p = \frac{B}{2R^2\sin\alpha} \, \, ,
\end{equation}
where $B$ is the bending stiffness and $R$ is the tube radius value.

In the presence of adhesion between a tip and a substrate, an adhesive
pressure ($p_a$) proportional to the surface energy $\gamma$ (by a
factor of $c/t$, with $c$ constant and $t$ thickness of the layer)
naturally emerges. Accordingly~\cite{pugno1}:
\begin{equation}
\label{pa}
\frac{p_a}{p}\approx \frac{c}{t} \, \frac{\gamma R^2}{B} \, .
\end{equation}
Similarly, considering a variation of pressure ($p_{AT}$) as
imposed by a variation of the bending stiffness due to atomistic
effects~\cite{ref6}:
\begin{equation}
\label{pqm}
\frac{p_{AT}}{p}\approx - \frac{\Delta B}{B} \, .
\end{equation}
Imposing: $p' = p+p_a+p_{AT}=\frac{B}{2R^2\sin\alpha '}$, where the
$\alpha '$ is the nanoscale contact angle, and assuming that the
surface energy $\gamma$ is scaled down, we can derive the following
correction:

\begin{equation}
\label{alfalinha}
\sin\alpha '=\frac{\sin\alpha}{1+c\frac{\gamma R^2}{tB}-
  \frac{\Delta B}{B}} \, .
\end{equation}

When the system is large and atomistic effects negligible, we have, as
expected, $\alpha=\alpha '$, while for a single layer with a
fictitiously vanishing surface energy, % $n \rightarrow \infty$),
we have:
\begin{equation}
  \label{sinalfalinha}
\sin\alpha '=\frac{\sin\alpha}{1-\frac{\Delta B}{B}} \, .
\end{equation}

Using the data presented in Figure 4 in association with the Equation
(\ref{alfalinha}), we can obtain another estimation of the correction
values of the bending stiffness variation of a single layer structure,
$\Delta B/B$}.  Through all these assumptions, our corrected model
fits very well the data for the angle $\alpha$ as a function of the
scaling factor $n$} as shown in Figure 4.  We then obtain an estimate
of a variaton of about 15\% on the bending stiffness, which are in
excellent agreement with the prediction of Zhang and
collaborators~\cite{ref6}, thus strongly suggesting that atomistic
effects of the single (non-continuous) atomic layer structure are responsible for the $\alpha$ discrepancies.

% \section{Conclusion}

In summary, we have demonstrated that the so-called ``universal''
behavior for the conformations of sheets confined inside cylindrical
tubes~\cite{ref7}, is violated at nanoscale. The origin of the
discrepancies between the macro and nanoscale models cannot only be
attributed to the relative importance of van der Waals forces, but
have also atomistic
contributions, as recently predicted by Zhang and
collaborators~\cite{ref6}.

%\begin{acknowledgments}
\section{Acknowledgments}
 This work was supported in part by the Brazilian Agencies CNPq and
  FAPESP. The authors thank the Center for Computational Engineering
  and Sciences at Unicamp for financial support through the
  FAPESP/CEPID Grant \#2013/08293-7. AFF also acknowledges support
  from FAPESP grant \#2012/10106-8. NMP is supported by the European
  Research Council, Grants Bihsnam and Replica2 as well as by the
  European Union, within the Graphene Flagship.

%\end{acknowledgments}


\end{document}